# Twist-grain boundary phase characterized by AFM technique

Vladimíra Novotná,* Lubor Lejček, and Ladislav Fekete

*Institute of Physics of the Czech Academy of Sciences, Na Slovance 2, CZ-182 21 Prague 8, Czech Republic*

e-mail for correspondence: novotna@fzu.cz

Abstract

Twist grain boundary (TGB) phases represent liquid crystalline systems with a regular array of defects. In our research, we studied a compound with a stable TGBC phase and established its structure by various experimental techniques. We observed the surface of the smectic film by AFM microscope and detected a periodic relief. We found that the displacement amplitude is few nanometers with the periodicity about 500 nm. Such periodicity is in accordance with the periodicity of the TGBC blocks' rotation estimated by polarizing microscopy. The surface modulation is explained by the deformation of TGBC structure, which is created on TGBC films. A simplified model interpreting the observed smectic surface displacement to occur as a consequence of rotating TGBC blocks inside the sample is proposed. TGBC blocks deform differently depending on their orientation with respect to the force acting by the tip of AFM microscope cantilever probing the smectic surface.

**Introduction**

Twist grain boundary (TGB) phases are an example of frustrated liquid crystalline mesophases, which is connected with molecular chirality [1-2]. The TGB phases mostly appear as an intermediate state between cholesteric (Ch) and smectic phases. They have been theoretically described by Renn and Lubensky [3-5] and later were proved experimentally by Goodby et al. [6]. The TGB phases represent an example of lamellar (smectic) phases with a regular defect structure consisting of parallel screw dislocations, which separate smectic blocks (grains). The layer normal of smectic grains rotates across each grain boundary and creates a helical superstructure with an axis parallel to the smectic layers. In the TGB phase, there is a competition between chiral forces, twist of the molecules and their tendency to pack into layers. These liquid crystalline phases exhibit specific structural features and their observations in polarized light microscope (POM) show variable textures. We are speaking about TGBA phase if the molecules in smectic blocks are parallel to the layer normal. In the case of the molecules tilted with respect to the layer normal, we can obtain TGBC phase and there is a variety of different modifications. For the first time, the TGBC phase was proved experimentally by Nguyen et al. [7] and later several kinds of TGBC phases with different local smectic structures have been described [8-10]. While the structure of the TGBA phase seems to be well understood, the character and properties of tilted variants (TGBC phases) are still a subject of intensive research [11-13].

During our research of chiral molecules with a lactic acid unit in the chiral chain, we have found several compounds with TGB mesophases [14-22]. We have studied and theoretically described the nucleation of filaments in the TGBA phase under different conditions [23-24]. Recently, we synthesized and studied mesomorphic behaviour of new lactic acid derivatives designed as nKDDL and found a presence of TGB phases persisting at room temperatures [25]. For one homologue 12KDDL with n=12 (dodecyl in the non-chiral alkyl chain), the cholesteric - TGBA – TGBC phase sequence was found on cooling from the isotropic phase. The optical and dielectric properties of nKDDL compounds were studied and the effect of an electric field analysed [25].

In this contribution, we investigate the homologue 12KDDL by AFM microscopy and other experimental studies to gain more information on the structure of the TGBC phase. By application of AFM technique, we succeeded to visualize a defect structure in details. Additionally, we propose a model to describe our experimental data. We take into consideration the description of a relative displacement on the sample surface during AFM measurements. We will interpret the AFM data using a simplified elastic model characterizing the influence of the probing apex of AFM tip on the liquid crystalline sample surface in the TGBC phase.

.

**Experimental**

Synthesis of the studied compound 12KDDL was presented previously [25] and the cholesteric (Ch)-TGBA-TGBC phase sequence was established on cooling from the isotropic phase (Iso). We established the following values: 83ºC for Iso-Ch, 73ºC for Ch-TGBA and 42ºC for TGBA-TGBC phase transition temperatures. The melting point (m.p.) was established to be m.p.=49ºC, the compound crystallized at 14ºC during the DSC measurements. Our observations under a polarizing optical microscope (POM) in the TGBA phase in planar geometry revealed features of the upper-temperature cholesteric phase, with typical oily-streak textures. In the TGBC phase, a stripe or grid texture was found (Figure 1).

To gain more information about the TGB structure of 12KDDL, we prepared a film from the melted material. At the room temperature, we studied its surface by AFM microscope and observed a complex relief, which is presented in Figures 2a and 2b for illustration. Atomic Force Microscopy (AFM) measurements were carried out at room temperature on an ambient AFM (Bruker, Dimension Icon) in Peak Force Tapping mode with ScanAsyst Air tips (Bruker; k=0.4 N/m; nominal tip radius 2 nm).

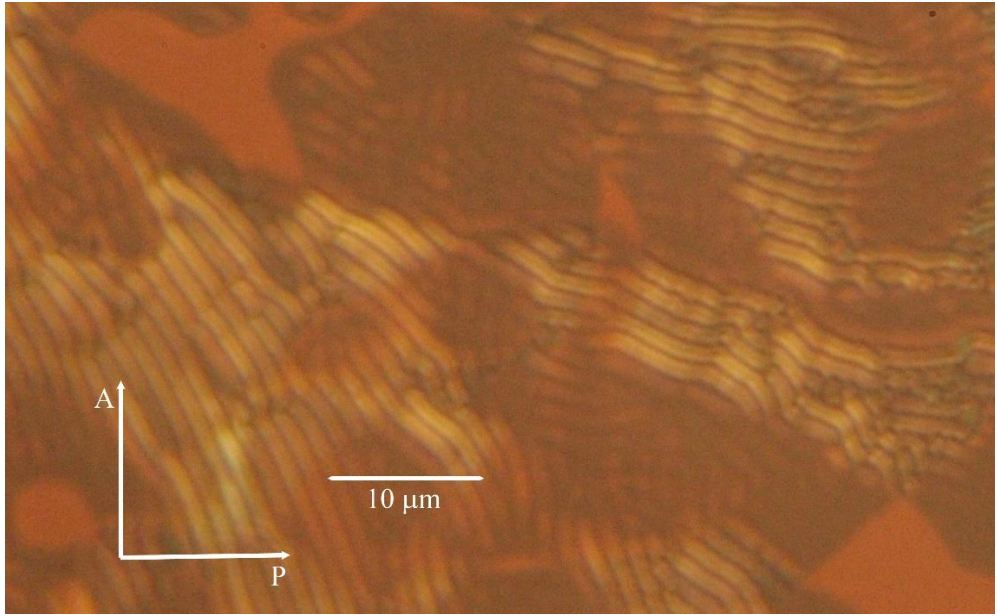

Figure 1.     Photo of the TGBC texture taken for the studied material at temperature T=30ºC in the polarizing microscope with crossed polarizer (P) and analyser (A) positions.

   The textures observed by AFM technique, which mark the surface relief of the sample, are in a form of the surface stripes. In Figure 2, we demonstrate two AFM figures with the depth profile of the surface amplitudes, which is in order of nanometres. We can analyse the profile of the studied sample in the direction perpendicular to these stripes precisely to obtain the plot of displacement over the surface. We expect that the revealed surface relief is created during the measurement as a result of the interaction between the sample elasticity and the force of AFM cantilever applying on the sample surface. To orient the system of coordinates with respect to a modulation, we present a schematic picture of an array of TGB blocks in the TGBA and TGBC phases in Figure 3. Such orientation will be necessary in the following theoretical description.

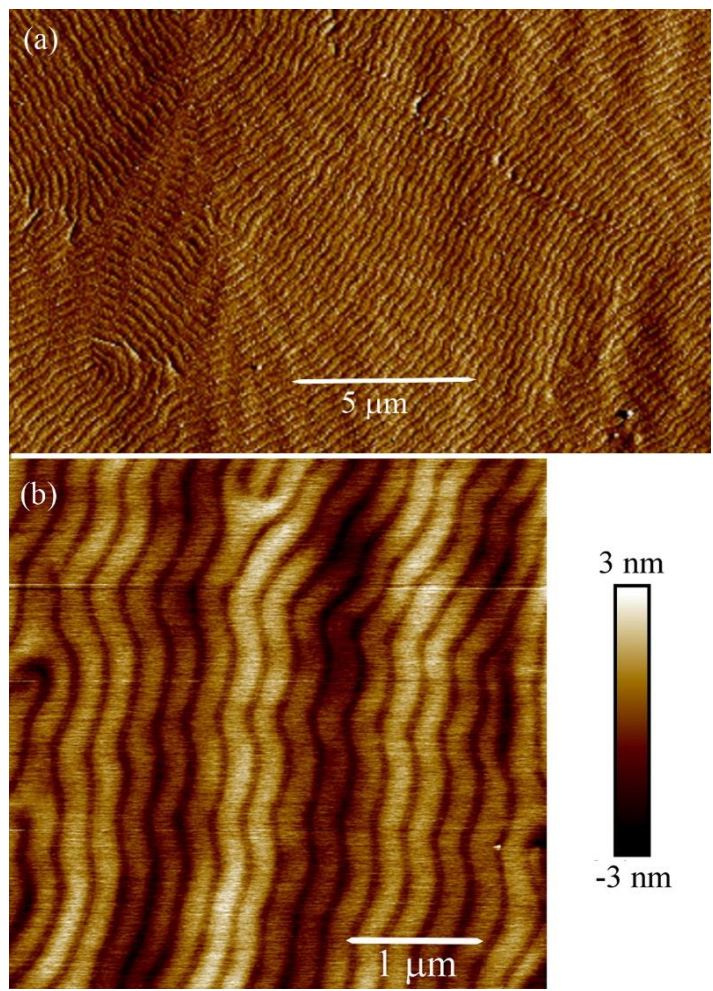

Figure 2. AFM pictures of the film surface at temperature about 30°C for 12KDDL sample. (a) Picture shows a global view and (b) a detailed character on the film surface. The intensity of colour corresponds to the surface relief described by relative displacement in nm. Measured topographies have 512x512 points resolution.

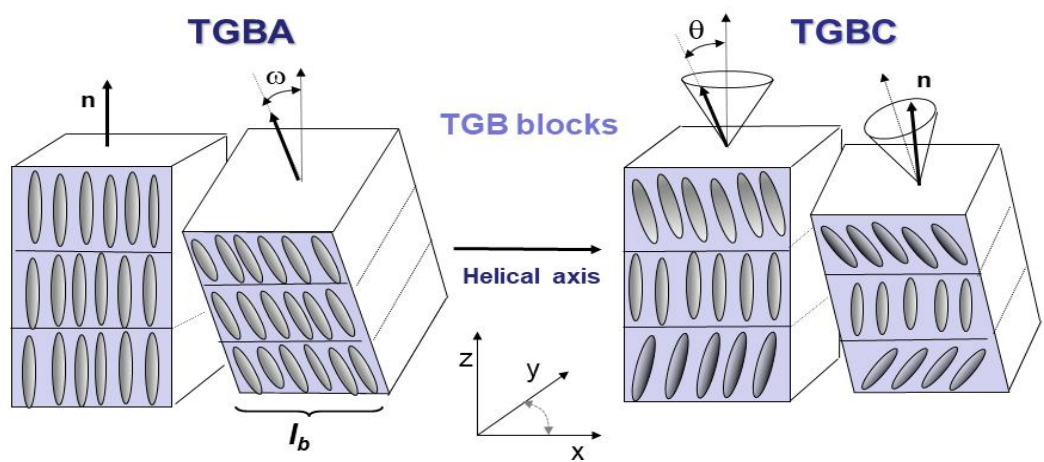

Figure 3. Coordinate axes and schematic organization of TGB blocks composed of molecular layers in TGBA and TGBC phases. Angle ω is the angle of the relative rotation of the neighbouring blocks with thickness $l_b$. In TGBC phase, vector **n** is the molecular direction in smectic C layers while θ is the molecular tilt in smectic C layers.

From AFM measurements, we can estimate that the period of the surface displacement is about 0.25 µm and a profile depth 3-5 nm. We expect that the surface displacement corresponds to the TGB structure inside the sample and not to the modulation of chiral smectic C director in TGBC structure. The periodicity of a modulation in the same range was observed by the absorption spectroscopy of the film, showing an anomaly in this spectral region which can be attributed to selective reflection. Unfortunately, a modulation of a defect array, seen by POM observations (Figure 1), cannot be properly analyzed due to the resolution of this technique. Therefore, we concentrated on AFM data, which are precise enough. In the next step, we will try to construct a model to explain the measured surface displacement using a description of TGB structure inside the sample and, at the end, to fit the experimental data we gained by AFM technique.

**Deformation of the TGB structure**

Herein, we will introduce a model to describe our observation of the surface displacement. As the sample is prepared as a film spread on the glass plate, we expect the molecules on the glass surface prefer homogeneous alignment. On contrary, on the film surface with a borderline with an air, we suppose a homeotropic molecular arrangement. Let us assume that the surface of the film consists of a narrow smectic layer of a thickness $d_o$. In this upper layer, as well as in the interior of the film, the layers are organized in a helical structure characteristic for TGB phases (Figure 3). We expect that in the central part the layers are relatively inclined to the glass surface and rotate around the axis denoted as *x*-axis (Figure 3). We can model the angle of layer inclination with respect to *y*-axis as $\phi(x)$, see Figure 4. This angle is a function of the variable *x* as it was presented previously [23]:

$$\phi(x) = \frac{\omega}{2} + \frac{\omega}{l_B} x. \qquad (1),$$

where $l_B$ is the dimension of TGB block along the axis *x*. Let *P* be the helical pitch along *x*-axis. Then the angle $\omega$ is the angle of a relative rotation of neighbouring TGB blocks (Figure 3). It can be determined as $\omega = 2\pi l_B/P$.

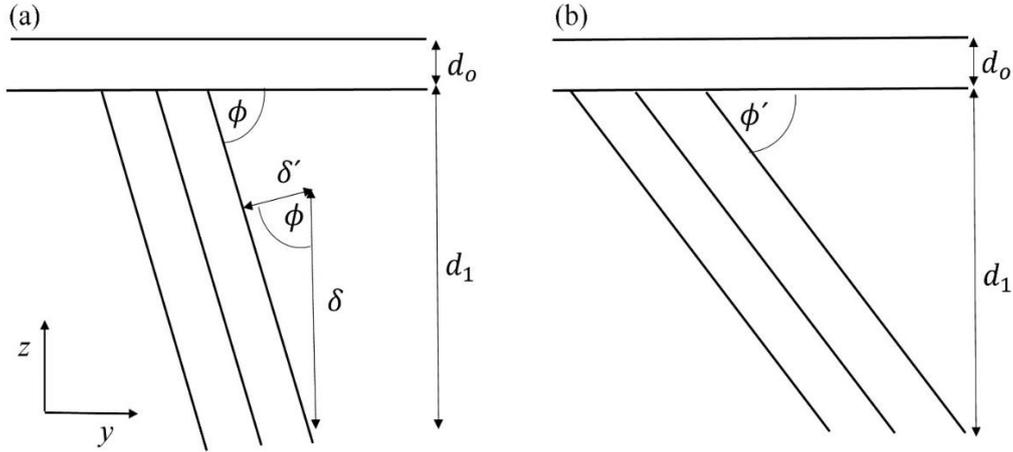

Figure 4. Scheme of the smectic layers orientation in (y,z)-plane of TGB film at a border with an air (upper part). Angle $\phi$ is the angle of inclination of the smectic layers in TGBA structure from *y*-axis and describes the rotation of TGBA blocks along the x-axis (Figure 3). (a) Displacement $\delta$ is induced by the microscope tip on the sample surface while $\delta'$ is the component of $\delta$ into the normal of the layers, inclined by an angle $\phi$. Parameter $d_o$ is the thickness of the smectic layers forming the sample surface, which is perpendicular to *z*-axis. The thickness $d_1$ describes this part of the sample where layers are organized in the helical order of TGB structure. (b) In the neighbouring TGB block oriented along x-axis the layers are inclined by angle $\phi' = \phi + \omega$.

We deal with the TGBC phase, in which smectic molecules in layers are tilted and create a helicoidal molecular arrangement with molecules rotating in the direction of layer normal. However, as AFM probing induces layer deformations, we neglect smectic C director structure for simplicity and we describe the smectic deformation using smectic A elastic approximation in our model. Therefore, we will describe the layer deformation of smectic liquid crystal by the free energy density $\rho f$ in the form:

$$\rho f = \frac{K}{2}\left(\frac{\partial^2 u}{\partial x^2} + \frac{\partial^2 u}{\partial y^2}\right)^2 + \frac{B}{2}\left(\frac{\partial u}{\partial z}\right)^2. \qquad (2)$$

*K* and *B* are the mean Frank elastic constants characterizing the layer curvature and compression modulus, respectively. The layer displacement in the direction of the layer normal is denoted as *u*. The free energy density yields the equilibrium equation:

$$\lambda^2 \Delta\Delta u = \frac{\partial^2 u}{\partial z^2}, \qquad (3)$$

where $\Delta = \frac{\partial^2}{\partial x^2} + \frac{\partial^2}{\partial y^2}$ and $\lambda = \sqrt{\frac{K}{B}}$. We propose a model solution of (3) to be isotropic in the plane $(x, y)$ in cylindrical coordinates:

$$u = \delta\left(\frac{2b_r^2}{r_o^2}\right)\left(\frac{r_o^2 - r^2}{2b_r^2} + \frac{z}{b_z}\right), \qquad (4)$$

with $r^2 = x^2 + y^2$. The solution (4) is defined for $r^2 \leq r_o^2$ and $0 \geq z \geq -d_z$. It means that (4) is non-zero inside of the volume $\pi r_o^2 d_z$ where $r_o^2 = d_x^2 + d_y^2$. Outside of this volume we take $u$=0. Parameters $d_x$, $d_y$ and $d_z$ are the dimensions of the mentioned volume in $x$-, $y$- and $z$- directions. Parameters $b_r$ and $b_z$ are the lengths over which the derivatives of the displacement $u$ change along radius $r$ and along $z$- direction, respectively. The volume $\pi r_o^2 d_z$ is situated under the sample surface at the position where the tip of AFM microscope cantilever is applied. As for boundary conditions of (4), we have $u$=0 at $r^2 = r_o^2$ and $z$=0. At $r$=0 it is $u = \delta$. The parameter $\delta$ is the amplitude of the displacement in $z$-direction induced by the AFM probing tip on the sample surface. From experimental data it follows that the order of $|\delta|$ corresponds to the smectic layer thickness value. On other surfaces of the volume $\pi r_o^2 d_z$, the displacement $u$ could have a jump-like character for about $\delta$ with respect to a non-deformed medium outside this volume. In this sense, the solution of (4) is an approximate description of the problem. Parameters will be determined by fitting $\delta$ to the experimental curve (Figure 5) in the next paragraph. The thickness $d_z$ is the sum $d_z = d_o + d_1$. We have already denoted the thickness $d_o$ as the thickness of the sample near its surface, where smectic layers are perpendicular to the direction of the force of AFM tip. The thickness $d_1$ describes the part of the sample where layers are organized in the helical order of TGB structure.

We can describe the elasticity of this part of the sample by the free energy density of (2) in the coordinate system $(x, y', z')$, which is rotated with respect to the coordinate system $(x, y, z)$ by an angle $\phi$ around the rotation $x$-axis. The displacement (4) will have the same form, however, in new coordinates. For simplicity, we will take parameters $b_x$ and $b_z$ the same in both coordinate systems. On the other hand, $\delta' = \delta \cos\phi$ (see Figure 4). The parameter $\delta'$ is the component of $\delta$ perpendicular to the smectic layer inclined from the $y$-axis by an angle $\theta$. There is also the component of $\delta$ along smectic layers. Nevertheless, this component is not taken into account as it does not contribute to the elasticity of smectic liquid crystal.

The total elastic free energy $F$ can be obtained using expressions (4) multiplied by volumes of corresponding volumes of considered structures:

$$F = (d_o \pi r_o^2) \frac{B}{2} \left(\frac{2b_r^2}{r_o^2}\right)^2 \left[\lambda^2 \left(\frac{2\delta}{b_r^2}\right)^2 + \left(\frac{\delta}{b_z}\right)^2\right] +$$

$$+ (d_1 \pi r_o^2) \frac{B}{2} \left(\frac{2b_r^2}{r_o^2}\right)^2 \left[\lambda^2 \left(\frac{2\delta}{b_r^2}\right)^2 + \left(\frac{\delta}{b_z}\right)^2\right] \cos^2 \phi \qquad (5).$$

Probing tip of AFM microscope acts on the sample surface by the force $-f_s$ per the surface $S$ of the microscope tip, which gives the surface stress $\sigma_s = -f_s/S$. The sign (-) means that the force is oriented in the opposite direction of $z$-axis.

The equilibrium of forces in our model is given by the relation:

$$-\frac{\partial F}{\partial \delta} + \sigma_s \pi r_o^2 = 0 \qquad (6).$$

Force $f_s$ is of the order of several pN but the surface of the microscope cantilever is difficult to determine. Let us define a non-dimensional force parameter $q = \left(\frac{|\sigma_s|}{B}\right)\left(\frac{r_o^2}{2b_r^2}\right)^2$, which will be later determined from the experimental dependence, see Figure 5. Using formula (5), the equilibrium equation (6) gives parameter $\delta$ in the form:

$$\delta = -\frac{qb_z^2 b_r^4}{(b_r^4+4\lambda^2 b_z^2)(d_o+d_1\cos^2\phi)} \qquad (7).$$

Expression (7) gives a relatively simple dependence of the displacement amplitude $\delta$ as the function of the angle of structure rotation along the rotation axis $x$. The change of amplitude $\delta$ in (7) is connected with the inclination $\phi$ of TGB grains with respect to the acting force $f_s$. When $\phi$ is small, force $f_s$ compresses smectic layers not only on the surface, but also in smectic blocs under the surface and $\delta$ is maximal. When $\phi$ approaches $\frac{\pi}{2}$, smectic layers in TGB blocks are not compressed and $\delta$ is minimal.

**Discussion**

The absolute values of displacement measured by AFM, as is demonstrated in Figure 4, are related to mean values of the surface position. In Figure 5, the experimental data measured by AFM technique are presented and the relative displacement of the sample surface was adjusted to be positive. Parameter $\delta$, together with other parameters in expression (7), can be fitted to the experimental data. As the differences of amplitude $\delta$ are important, we modify (7) by a constant, which shifts $\delta$ to obtain $\delta \geq 0$. Then we take

$$\delta = \frac{qb_z^2 b_r^4}{(b_r^4+4\lambda^2 b_z^2)}\left(\frac{1}{d_o} - \frac{1}{(d_o+d_1\cos^2\phi)}\right). \qquad (8)$$

For the purpose of fitting, we will modify the function $\phi(x)$ (1) as

$$\phi(x) = \frac{2\pi}{P}x - \frac{\pi}{2} + \phi_o . \qquad (9)$$

From (1) one can see that $x$ should change by jumps of the length $l_B$. For simplicity, we use (9) in expression (8) continuously. Parameter $\phi_o$ together with $-\frac{\pi}{2}$ in (9) is the phase shift of the angle $\phi(x)$, which permits us to adjust the function $\delta$ given by (8) to measured data. The reason is that we do not know the exact position of the coordinate origin. For the fit, we applied the estimation $\lambda$=4.5 nm as the starting value, which corresponds to the molecular length of 12KDDL at the temperature about 5 K below the temperature transition from TGBA to TGBC [25].

The fit of $\delta$ to experimental data is presented in Figure 5. Maxima of $\delta$ correspond to $\phi$ giving $cos^2\phi = 1$. The fitting gives the following values of parameters: the pitch of TGB structure $P$=0.50 μm, $b_r$ =11.3 nm, $b_z$ =15.5 nm, $d_o$=4.2 nm, $d_1$=11.6 nm and $\lambda$=4.61 nm. The parameter of the phase shift $\theta_o$ = -2.29. Finally, the non-dimensional force parameter $q$ was determined as $q$ =0.17.

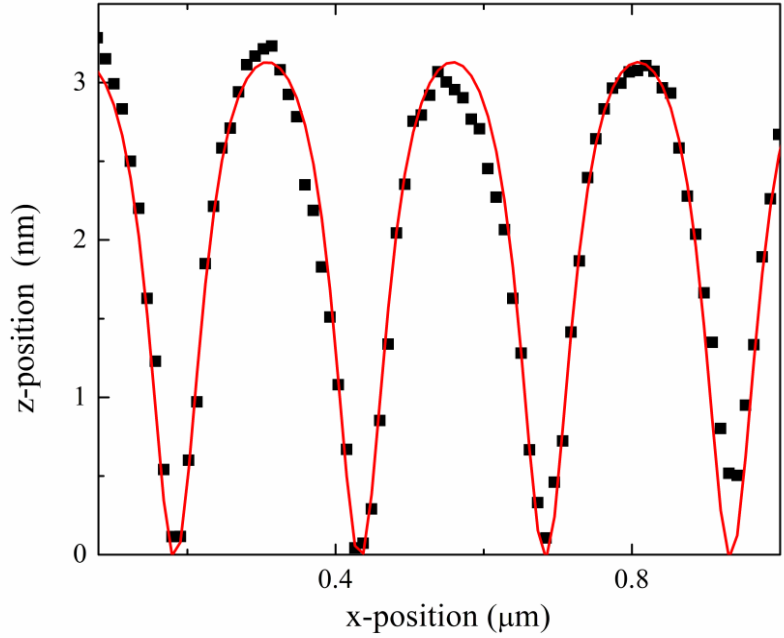

Figure 5.    Experimental values of the surface displacement profile (along z-axis) versus a relative *x*-position at the surface of the film (in μm), taken from AFM measurements (full squares). The full line (in red colour) is a fit of $\delta$ to the expression (9).

Let us discuss the values of parameters we obtained by fitting the AFM data. Let us point out that the pitch length, $P$, is twice this periodicity value, and similar values were found for another lactic acid derivative exhibiting the TGBA phase, which was studied previously [21]. The periodicity is also comparable to other TGB phases, which were presented previously for different materials, see eg. [1-3]. Using our parameters, we can calculate the volume $\pi r_o^2 d_z$. The length $d_z$ can be written as $d_z = d_o + d_1 \approx 14.5$ nm. This value is nearly equal to $b_z = 15.2$ nm. Then we can expect that $d_x \approx d_y \approx b_r = 11$ nm. Therefore, $\pi r_o^2 = \pi(d_x^2 + d_y^2) \approx 2\pi b_r^2$. We also suppose that $d_x \approx l_B$. In Ref. [1], the length $l_B$ for another type of the TGB material was estimated to be 18.3 nm, so we consider such a value as reasonable. The parameter $q = \left(\frac{|f_s|}{BS}\right)\left(\frac{r_o^2}{2b_r^2}\right)^2 = 0.17$ gives the relation among force $f_s$ action on the surface $S$ of the sample and the compression modulus $B$. The value of $B$ is the function of the temperature and it can be roughly estimated from the value $\lambda = 4.61$ nm we obtained from fitting procedure. If we utilize value $K \sim 10^{-11}$ N, which is presented in the literature [1], we obtain $B \sim 4.7 \times 10^5$ J/m$^3$. This value is smaller than in smectic A phase, which reflects the fact that we deal with a tilted smectic C structure, in which $B$ is smaller than in smectic A phases.

**Conclusions**

In summary, we have studied the films of the lactic acid derivative 12KDDL in the TGBC phase by AFM technique. Then we have analysed our observations and tried to describe the structural parameters from our experimental data. Based on a detailed analysis of AFM films, we have prepared a simplified theoretical model, which reflects the topology of the TGB phase as well as its elastic properties. In our model we consider the TGB structure to be composed of inclined layers enveloped by a smectic layer, which ensured homeotropic anchoring at a boarder with an air. This consideration gives us a reasonable background for fitting the experimental data. The model assumes an equilibrium between the force of the microscope tip acting and probing the sample surface and the helicoidal TGB structure. The deformation of TGB structure depends on the orientation of smectic layers in TGB structure with respect to the orientation of acting force, which leads to the change in a relative displacement $\delta$. Therefore, it is important to note that the microscope probing the sample surface reveals also the TGB structure under the sample surface. The fit of the measured surface displacement is in concordance with the displacement obtained from our theoretical model. The observed stripe periodicity detected by AFM corresponds to the periodicity of TGB structure inside the sample and offers us an important information.

The values of the surface displacement amplitude $\delta$, which we fitted to our experimental data, are in agreements with other microscopic and spectral observations on 12KDDL. Unfortunately, we are not able to establish the periodicity observed in the polarizing microscope (Figure 1) with a better resolution. Additionally, the spectroscopic measurements on the film detected the anomaly in rather broad range 500-600 nm and it is also difficult to establish the periodicity value exactly based on this method. We can summarize that AFM technique is a very powerful technique and gives us very precise information about the pitch value. The first attempt to apply AFM measurements to establish the periodicity of a cholesteric phase with a short pitch-length was done by Škarabot et al. [26]. Nevertheless, to the best of our knowledge, TGB phases have not yet been characterized by this technique.

**Disclosure statement**

No potential conflict of interest was reported by the authors.

**Funding**

This work was supported by the Czech Ministry of Education, Youth and Sports [project No. LTC19051] and Operational Programme Research, Development and Education financed by European Structural and Investment Funds [project SOLID21-CZ.02.1.01/0.0/0.0/16_019/0000760]. The authors would like to acknowledge the contribution of the COST Action CA17139EU "European Topology Interdisciplinary Action".